\documentstyle[12pt]{article}
\begin{document}
\begin{flushright}
{\bf Preprint SSU-HEP-00/09\\
Samara State University}
\end{flushright}
\vspace{30mm}
\begin{center}
{\bf MAGNETIC MOMENT OF THE TWO-PARTICLE\\
BOUND STATE IN QUANTUM ELECTRODYNAMICS}\footnote{
Talk presented at Nuclear Physics Department Conference
"Physics of Fundamental Interactions" Russian Academy of
Sciences, ITEP, Moscow, 27 November - 1 December 2000}\\

\vspace{4mm}

R.N.~Faustov \\Scientific Council "Cybernetics" RAS\\
117333, Moscow, Vavilov, 40, Russia,\\
A.P.~Martynenko\\ Department of Theoretical Physics, Samara State University,\\
443011, Samara, Pavlov, 1, Russia
\end{center}

\begin{abstract}
We have formulated the quasipotential method for the calculation
of the relativistic and radiative corrections to the magnetic
moment of the two-particle bound state in the case of particles
with arbitrary spin. It is shown that the g-factors of bound particles
contain $O(\alpha^2)$ terms depending on the particle spin. Numerical
values for the g-factors of the electron in the hydrogen atom and
deuterium are obtained.
\end{abstract}

\immediate\write16{<<WARNING: LINEDRAW macros work with emTeX-dvivers
                    and other drivers supporting emTeX \special's
                    (dviscr, dvihplj, dvidot, dvips, dviwin, etc.) >>}

\newdimen\Lengthunit       \Lengthunit  = 1.5cm
\newcount\Nhalfperiods     \Nhalfperiods= 9
\newcount\magnitude        \magnitude = 1000

\catcode`\*=11
\newdimen\L*   \newdimen\d*   \newdimen\d**
\newdimen\dm*  \newdimen\dd*  \newdimen\dt*
\newdimen\a*   \newdimen\b*   \newdimen\c*
\newdimen\a**  \newdimen\b**
\newdimen\xL*  \newdimen\yL*
\newdimen\rx*  \newdimen\ry*
\newdimen\tmp* \newdimen\linwid*

\newcount\k*   \newcount\l*   \newcount\m*
\newcount\k**  \newcount\l**  \newcount\m**
\newcount\n*   \newcount\dn*  \newcount\r*
\newcount\N*   \newcount\*one \newcount\*two  \*one=1 \*two=2
\newcount\*ths \*ths=1000
\newcount\angle*  \newcount\q*  \newcount\q**
\newcount\angle** \angle**=0
\newcount\sc*     \sc*=0

\newtoks\cos*  \cos*={1}
\newtoks\sin*  \sin*={0}

\catcode`\[=13

\def\rotate(#1){\advance\angle**#1\angle*=\angle**
\q**=\angle*\ifnum\q**<0\q**=-\q**\fi
\ifnum\q**>360\q*=\angle*\divide\q*360\multiply\q*360\advance\angle*-\q*\fi
\ifnum\angle*<0\advance\angle*360\fi\q**=\angle*\divide\q**90\q**=\q**
\def\sgcos*{+}\def\sgsin*{+}\relax
\ifcase\q**\or
 \def\sgcos*{-}\def\sgsin*{+}\or
 \def\sgcos*{-}\def\sgsin*{-}\or
 \def\sgcos*{+}\def\sgsin*{-}\else\fi
\q*=\q**
\multiply\q*90\advance\angle*-\q*
\ifnum\angle*>45\sc*=1\angle*=-\angle*\advance\angle*90\else\sc*=0\fi
\def[##1,##2]{\ifnum\sc*=0\relax
\edef\cs*{\sgcos*.##1}\edef\sn*{\sgsin*.##2}\ifcase\q**\or
 \edef\cs*{\sgcos*.##2}\edef\sn*{\sgsin*.##1}\or
 \edef\cs*{\sgcos*.##1}\edef\sn*{\sgsin*.##2}\or
 \edef\cs*{\sgcos*.##2}\edef\sn*{\sgsin*.##1}\else\fi\else
\edef\cs*{\sgcos*.##2}\edef\sn*{\sgsin*.##1}\ifcase\q**\or
 \edef\cs*{\sgcos*.##1}\edef\sn*{\sgsin*.##2}\or
 \edef\cs*{\sgcos*.##2}\edef\sn*{\sgsin*.##1}\or
 \edef\cs*{\sgcos*.##1}\edef\sn*{\sgsin*.##2}\else\fi\fi
\cos*={\cs*}\sin*={\sn*}\global\edef\gcos*{\cs*}\global\edef\gsin*{\sn*}}\relax
\ifcase\angle*[9999,0]\or
[999,017]\or[999,034]\or[998,052]\or[997,069]\or[996,087]\or
[994,104]\or[992,121]\or[990,139]\or[987,156]\or[984,173]\or
[981,190]\or[978,207]\or[974,224]\or[970,241]\or[965,258]\or
[961,275]\or[956,292]\or[951,309]\or[945,325]\or[939,342]\or
[933,358]\or[927,374]\or[920,390]\or[913,406]\or[906,422]\or
[898,438]\or[891,453]\or[882,469]\or[874,484]\or[866,499]\or
[857,515]\or[848,529]\or[838,544]\or[829,559]\or[819,573]\or
[809,587]\or[798,601]\or[788,615]\or[777,629]\or[766,642]\or
[754,656]\or[743,669]\or[731,681]\or[719,694]\or[707,707]\or
\else[9999,0]\fi}

\catcode`\[=12

\def\GRAPH(hsize=#1)#2{\hbox to #1\Lengthunit{#2\hss}}

\def\Linewidth#1{\global\linwid*=#1\relax
\global\divide\linwid*10\global\multiply\linwid*\mag
\global\divide\linwid*100\special{em:linewidth \the\linwid*}}

\Linewidth{.4pt}
\def\sm*{\special{em:moveto}}
\def\sl*{\special{em:lineto}}
\let\moveto=\sm*
\let\lineto=\sl*
\newbox\spm*   \newbox\spl*
\setbox\spm*\hbox{\sm*}
\setbox\spl*\hbox{\sl*}

\def\mov#1(#2,#3)#4{\rlap{\L*=#1\Lengthunit
\xL*=#2\L* \yL*=#3\L*
\xL*=\xscale\xL* \yL*=\yscale\yL*
\rx* \the\cos*\xL* \tmp* \the\sin*\yL* \advance\rx*-\tmp*
\ry* \the\cos*\yL* \tmp* \the\sin*\xL* \advance\ry*\tmp*
\kern\rx*\raise\ry*\hbox{#4}}}

\def\rmov*(#1,#2)#3{\rlap{\xL*=#1\yL*=#2\relax
\rx* \the\cos*\xL* \tmp* \the\sin*\yL* \advance\rx*-\tmp*
\ry* \the\cos*\yL* \tmp* \the\sin*\xL* \advance\ry*\tmp*
\kern\rx*\raise\ry*\hbox{#3}}}

\def\lin#1(#2,#3){\rlap{\sm*\mov#1(#2,#3){\sl*}}}

\def\arr*(#1,#2,#3){\rmov*(#1\dd*,#1\dt*){\sm*
\rmov*(#2\dd*,#2\dt*){\rmov*(#3\dt*,-#3\dd*){\sl*}}\sm*
\rmov*(#2\dd*,#2\dt*){\rmov*(-#3\dt*,#3\dd*){\sl*}}}}

\def\arrow#1(#2,#3){\rlap{\lin#1(#2,#3)\mov#1(#2,#3){\relax
\d**=-.012\Lengthunit\dd*=#2\d**\dt*=#3\d**
\arr*(1,10,4)\arr*(3,8,4)\arr*(4.8,4.2,3)}}}

\def\arrlin#1(#2,#3){\rlap{\L*=#1\Lengthunit\L*=.5\L*
\lin#1(#2,#3)\rmov*(#2\L*,#3\L*){\arrow.1(#2,#3)}}}

\def\dasharrow#1(#2,#3){\rlap{{\Lengthunit=0.9\Lengthunit
\dashlin#1(#2,#3)\mov#1(#2,#3){\sm*}}\mov#1(#2,#3){\sl*
\d**=-.012\Lengthunit\dd*=#2\d**\dt*=#3\d**
\arr*(1,10,4)\arr*(3,8,4)\arr*(4.8,4.2,3)}}}

\def\clap#1{\hbox to 0pt{\hss #1\hss}}

\def\ind(#1,#2)#3{\rlap{\L*=.1\Lengthunit
\xL*=#1\L* \yL*=#2\L*
\rx* \the\cos*\xL* \tmp* \the\sin*\yL* \advance\rx*-\tmp*
\ry* \the\cos*\yL* \tmp* \the\sin*\xL* \advance\ry*\tmp*
\kern\rx*\raise\ry*\hbox{\lower2pt\clap{$#3$}}}}

\def\sh*(#1,#2)#3{\rlap{\dm*=\the\n*\d**
\xL*=\xscale\dm* \yL*=\yscale\dm* \xL*=#1\xL* \yL*=#2\yL*
\rx* \the\cos*\xL* \tmp* \the\sin*\yL* \advance\rx*-\tmp*
\ry* \the\cos*\yL* \tmp* \the\sin*\xL* \advance\ry*\tmp*
\kern\rx*\raise\ry*\hbox{#3}}}

\def\calcnum*#1(#2,#3){\a*=1000sp\b*=1000sp\a*=#2\a*\b*=#3\b*
\ifdim\a*<0pt\a*-\a*\fi\ifdim\b*<0pt\b*-\b*\fi
\ifdim\a*>\b*\c*=.96\a*\advance\c*.4\b*
\else\c*=.96\b*\advance\c*.4\a*\fi
\k*\a*\multiply\k*\k*\l*\b*\multiply\l*\l*
\m*\k*\advance\m*\l*\n*\c*\r*\n*\multiply\n*\n*
\dn*\m*\advance\dn*-\n*\divide\dn*2\divide\dn*\r*
\advance\r*\dn*
\c*=\the\Nhalfperiods5sp\c*=#1\c*\ifdim\c*<0pt\c*-\c*\fi
\multiply\c*\r*\N*\c*\divide\N*10000}

\def\dashlin#1(#2,#3){\rlap{\calcnum*#1(#2,#3)\relax
\d**=#1\Lengthunit\ifdim\d**<0pt\d**-\d**\fi
\divide\N*2\multiply\N*2\advance\N*\*one
\divide\d**\N*\sm*\n*\*one\sh*(#2,#3){\sl*}\loop
\advance\n*\*one\sh*(#2,#3){\sm*}\advance\n*\*one
\sh*(#2,#3){\sl*}\ifnum\n*<\N*\repeat}}

\def\dashdotlin#1(#2,#3){\rlap{\calcnum*#1(#2,#3)\relax
\d**=#1\Lengthunit\ifdim\d**<0pt\d**-\d**\fi
\divide\N*2\multiply\N*2\advance\N*1\multiply\N*2\relax
\divide\d**\N*\sm*\n*\*two\sh*(#2,#3){\sl*}\loop
\advance\n*\*one\sh*(#2,#3){\kern-1.48pt\lower.5pt\hbox{\rm.}}\relax
\advance\n*\*one\sh*(#2,#3){\sm*}\advance\n*\*two
\sh*(#2,#3){\sl*}\ifnum\n*<\N*\repeat}}

\def\shl*(#1,#2)#3{\kern#1#3\lower#2#3\hbox{\unhcopy\spl*}}

\def\trianglin#1(#2,#3){\rlap{\toks0={#2}\toks1={#3}\calcnum*#1(#2,#3)\relax
\dd*=.57\Lengthunit\dd*=#1\dd*\divide\dd*\N*
\divide\dd*\*ths \multiply\dd*\magnitude
\d**=#1\Lengthunit\ifdim\d**<0pt\d**-\d**\fi
\multiply\N*2\divide\d**\N*\sm*\n*\*one\loop
\shl**{\dd*}\dd*-\dd*\advance\n*2\relax
\ifnum\n*<\N*\repeat\n*\N*\shl**{0pt}}}

\def\wavelin#1(#2,#3){\rlap{\toks0={#2}\toks1={#3}\calcnum*#1(#2,#3)\relax
\dd*=.23\Lengthunit\dd*=#1\dd*\divide\dd*\N*
\divide\dd*\*ths \multiply\dd*\magnitude
\d**=#1\Lengthunit\ifdim\d**<0pt\d**-\d**\fi
\multiply\N*4\divide\d**\N*\sm*\n*\*one\loop
\shl**{\dd*}\dt*=1.3\dd*\advance\n*\*one
\shl**{\dt*}\advance\n*\*one
\shl**{\dd*}\advance\n*\*two
\dd*-\dd*\ifnum\n*<\N*\repeat\n*\N*\shl**{0pt}}}

\def\w*lin(#1,#2){\rlap{\toks0={#1}\toks1={#2}\d**=\Lengthunit\dd*=-.12\d**
\divide\dd*\*ths \multiply\dd*\magnitude
\N*8\divide\d**\N*\sm*\n*\*one\loop
\shl**{\dd*}\dt*=1.3\dd*\advance\n*\*one
\shl**{\dt*}\advance\n*\*one
\shl**{\dd*}\advance\n*\*one
\shl**{0pt}\dd*-\dd*\advance\n*1\ifnum\n*<\N*\repeat}}

\def\l*arc(#1,#2)[#3][#4]{\rlap{\toks0={#1}\toks1={#2}\d**=\Lengthunit
\dd*=#3.037\d**\dd*=#4\dd*\dt*=#3.049\d**\dt*=#4\dt*\ifdim\d**>10mm\relax
\d**=.25\d**\n*\*one\shl**{-\dd*}\n*\*two\shl**{-\dt*}\n*3\relax
\shl**{-\dd*}\n*4\relax\shl**{0pt}\else
\ifdim\d**>5mm\d**=.5\d**\n*\*one\shl**{-\dt*}\n*\*two
\shl**{0pt}\else\n*\*one\shl**{0pt}\fi\fi}}

\def\d*arc(#1,#2)[#3][#4]{\rlap{\toks0={#1}\toks1={#2}\d**=\Lengthunit
\dd*=#3.037\d**\dd*=#4\dd*\d**=.25\d**\sm*\n*\*one\shl**{-\dd*}\relax
\n*3\relax\sh*(#1,#2){\xL*=\xscale\dd*\yL*=\yscale\dd*
\kern#2\xL*\lower#1\yL*\hbox{\sm*}}\n*4\relax\shl**{0pt}}}

\def\shl**#1{\c*=\the\n*\d**\d*=#1\relax
\a*=\the\toks0\c*\b*=\the\toks1\d*\advance\a*-\b*
\b*=\the\toks1\c*\d*=\the\toks0\d*\advance\b*\d*
\a*=\xscale\a*\b*=\yscale\b*
\rx* \the\cos*\a* \tmp* \the\sin*\b* \advance\rx*-\tmp*
\ry* \the\cos*\b* \tmp* \the\sin*\a* \advance\ry*\tmp*
\raise\ry*\rlap{\kern\rx*\unhcopy\spl*}}

\def\wlin*#1(#2,#3)[#4]{\rlap{\toks0={#2}\toks1={#3}\relax
\c*=#1\l*\c*\c*=.01\Lengthunit\m*\c*\divide\l*\m*
\c*=\the\Nhalfperiods5sp\multiply\c*\l*\N*\c*\divide\N*\*ths
\divide\N*2\multiply\N*2\advance\N*\*one
\dd*=.002\Lengthunit\dd*=#4\dd*\multiply\dd*\l*\divide\dd*\N*
\divide\dd*\*ths \multiply\dd*\magnitude
\d**=#1\multiply\N*4\divide\d**\N*\sm*\n*\*one\loop
\shl**{\dd*}\dt*=1.3\dd*\advance\n*\*one
\shl**{\dt*}\advance\n*\*one
\shl**{\dd*}\advance\n*\*two
\dd*-\dd*\ifnum\n*<\N*\repeat\n*\N*\shl**{0pt}}}

\def\wavebox#1{\setbox0\hbox{#1}\relax
\a*=\wd0\advance\a*14pt\b*=\ht0\advance\b*\dp0\advance\b*14pt\relax
\hbox{\kern9pt\relax
\rmov*(0pt,\ht0){\rmov*(-7pt,7pt){\wlin*\a*(1,0)[+]\wlin*\b*(0,-1)[-]}}\relax
\rmov*(\wd0,-\dp0){\rmov*(7pt,-7pt){\wlin*\a*(-1,0)[+]\wlin*\b*(0,1)[-]}}\relax
\box0\kern9pt}}

\def\rectangle#1(#2,#3){\relax
\lin#1(#2,0)\lin#1(0,#3)\mov#1(0,#3){\lin#1(#2,0)}\mov#1(#2,0){\lin#1(0,#3)}}

\def\dashrectangle#1(#2,#3){\dashlin#1(#2,0)\dashlin#1(0,#3)\relax
\mov#1(0,#3){\dashlin#1(#2,0)}\mov#1(#2,0){\dashlin#1(0,#3)}}

\def\waverectangle#1(#2,#3){\L*=#1\Lengthunit\a*=#2\L*\b*=#3\L*
\ifdim\a*<0pt\a*-\a*\def\x*{-1}\else\def\x*{1}\fi
\ifdim\b*<0pt\b*-\b*\def\y*{-1}\else\def\y*{1}\fi
\wlin*\a*(\x*,0)[-]\wlin*\b*(0,\y*)[+]\relax
\mov#1(0,#3){\wlin*\a*(\x*,0)[+]}\mov#1(#2,0){\wlin*\b*(0,\y*)[-]}}

\def\calcparab*{\ifnum\n*>\m*\k*\N*\advance\k*-\n*\else\k*\n*\fi
\a*=\the\k* sp\a*=10\a*\b*\dm*\advance\b*-\a*\k*\b*
\a*=\the\*ths\b*\divide\a*\l*\multiply\a*\k*
\divide\a*\l*\k*\*ths\r*\a*\advance\k*-\r*\dt*=\the\k*\L*}

\def\arcto#1(#2,#3)[#4]{\rlap{\toks0={#2}\toks1={#3}\calcnum*#1(#2,#3)\relax
\dm*=135sp\dm*=#1\dm*\d**=#1\Lengthunit\ifdim\dm*<0pt\dm*-\dm*\fi
\multiply\dm*\r*\a*=.3\dm*\a*=#4\a*\ifdim\a*<0pt\a*-\a*\fi
\advance\dm*\a*\N*\dm*\divide\N*10000\relax
\divide\N*2\multiply\N*2\advance\N*\*one
\L*=-.25\d**\L*=#4\L*\divide\d**\N*\divide\L*\*ths
\m*\N*\divide\m*2\dm*=\the\m*5sp\l*\dm*\sm*\n*\*one\loop
\calcparab*\shl**{-\dt*}\advance\n*1\ifnum\n*<\N*\repeat}}

\def\arrarcto#1(#2,#3)[#4]{\L*=#1\Lengthunit\L*=.54\L*
\arcto#1(#2,#3)[#4]\rmov*(#2\L*,#3\L*){\d*=.457\L*\d*=#4\d*\d**-\d*
\rmov*(#3\d**,#2\d*){\arrow.02(#2,#3)}}}

\def\dasharcto#1(#2,#3)[#4]{\rlap{\toks0={#2}\toks1={#3}\relax
\calcnum*#1(#2,#3)\dm*=\the\N*5sp\a*=.3\dm*\a*=#4\a*\ifdim\a*<0pt\a*-\a*\fi
\advance\dm*\a*\N*\dm*
\divide\N*20\multiply\N*2\advance\N*1\d**=#1\Lengthunit
\L*=-.25\d**\L*=#4\L*\divide\d**\N*\divide\L*\*ths
\m*\N*\divide\m*2\dm*=\the\m*5sp\l*\dm*
\sm*\n*\*one\loop\calcparab*
\shl**{-\dt*}\advance\n*1\ifnum\n*>\N*\else\calcparab*
\sh*(#2,#3){\xL*=#3\dt* \yL*=#2\dt*
\rx* \the\cos*\xL* \tmp* \the\sin*\yL* \advance\rx*\tmp*
\ry* \the\cos*\yL* \tmp* \the\sin*\xL* \advance\ry*-\tmp*
\kern\rx*\lower\ry*\hbox{\sm*}}\fi
\advance\n*1\ifnum\n*<\N*\repeat}}

\def\*shl*#1{\c*=\the\n*\d**\advance\c*#1\a**\d*\dt*\advance\d*#1\b**
\a*=\the\toks0\c*\b*=\the\toks1\d*\advance\a*-\b*
\b*=\the\toks1\c*\d*=\the\toks0\d*\advance\b*\d*
\rx* \the\cos*\a* \tmp* \the\sin*\b* \advance\rx*-\tmp*
\ry* \the\cos*\b* \tmp* \the\sin*\a* \advance\ry*\tmp*
\raise\ry*\rlap{\kern\rx*\unhcopy\spl*}}

\def\calcnormal*#1{\b**=10000sp\a**\b**\k*\n*\advance\k*-\m*
\multiply\a**\k*\divide\a**\m*\a**=#1\a**\ifdim\a**<0pt\a**-\a**\fi
\ifdim\a**>\b**\d*=.96\a**\advance\d*.4\b**
\else\d*=.96\b**\advance\d*.4\a**\fi
\d*=.01\d*\r*\d*\divide\a**\r*\divide\b**\r*
\ifnum\k*<0\a**-\a**\fi\d*=#1\d*\ifdim\d*<0pt\b**-\b**\fi
\k*\a**\a**=\the\k*\dd*\k*\b**\b**=\the\k*\dd*}

\def\wavearcto#1(#2,#3)[#4]{\rlap{\toks0={#2}\toks1={#3}\relax
\calcnum*#1(#2,#3)\c*=\the\N*5sp\a*=.4\c*\a*=#4\a*\ifdim\a*<0pt\a*-\a*\fi
\advance\c*\a*\N*\c*\divide\N*20\multiply\N*2\advance\N*-1\multiply\N*4\relax
\d**=#1\Lengthunit\dd*=.012\d**
\divide\dd*\*ths \multiply\dd*\magnitude
\ifdim\d**<0pt\d**-\d**\fi\L*=.25\d**
\divide\d**\N*\divide\dd*\N*\L*=#4\L*\divide\L*\*ths
\m*\N*\divide\m*2\dm*=\the\m*0sp\l*\dm*
\sm*\n*\*one\loop\calcnormal*{#4}\calcparab*
\*shl*{1}\advance\n*\*one\calcparab*
\*shl*{1.3}\advance\n*\*one\calcparab*
\*shl*{1}\advance\n*2\dd*-\dd*\ifnum\n*<\N*\repeat\n*\N*\shl**{0pt}}}

\def\triangarcto#1(#2,#3)[#4]{\rlap{\toks0={#2}\toks1={#3}\relax
\calcnum*#1(#2,#3)\c*=\the\N*5sp\a*=.4\c*\a*=#4\a*\ifdim\a*<0pt\a*-\a*\fi
\advance\c*\a*\N*\c*\divide\N*20\multiply\N*2\advance\N*-1\multiply\N*2\relax
\d**=#1\Lengthunit\dd*=.012\d**
\divide\dd*\*ths \multiply\dd*\magnitude
\ifdim\d**<0pt\d**-\d**\fi\L*=.25\d**
\divide\d**\N*\divide\dd*\N*\L*=#4\L*\divide\L*\*ths
\m*\N*\divide\m*2\dm*=\the\m*0sp\l*\dm*
\sm*\n*\*one\loop\calcnormal*{#4}\calcparab*
\*shl*{1}\advance\n*2\dd*-\dd*\ifnum\n*<\N*\repeat\n*\N*\shl**{0pt}}}

\def\hr*#1{\L*=\xscale\Lengthunit\ifnum
\angle**=0\clap{\vrule width#1\L* height.1pt}\else
\L*=#1\L*\L*=.5\L*\rmov*(-\L*,0pt){\sm*}\rmov*(\L*,0pt){\sl*}\fi}

\def\shade#1[#2]{\rlap{\Lengthunit=#1\Lengthunit
\special{em:linewidth .001pt}\relax
\mov(0,#2.05){\hr*{.994}}\mov(0,#2.1){\hr*{.980}}\relax
\mov(0,#2.15){\hr*{.953}}\mov(0,#2.2){\hr*{.916}}\relax
\mov(0,#2.25){\hr*{.867}}\mov(0,#2.3){\hr*{.798}}\relax
\mov(0,#2.35){\hr*{.715}}\mov(0,#2.4){\hr*{.603}}\relax
\mov(0,#2.45){\hr*{.435}}\special{em:linewidth \the\linwid*}}}

\def\dshade#1[#2]{\rlap{\special{em:linewidth .001pt}\relax
\Lengthunit=#1\Lengthunit\if#2-\def\t*{+}\else\def\t*{-}\fi
\mov(0,\t*.025){\relax
\mov(0,#2.05){\hr*{.995}}\mov(0,#2.1){\hr*{.988}}\relax
\mov(0,#2.15){\hr*{.969}}\mov(0,#2.2){\hr*{.937}}\relax
\mov(0,#2.25){\hr*{.893}}\mov(0,#2.3){\hr*{.836}}\relax
\mov(0,#2.35){\hr*{.760}}\mov(0,#2.4){\hr*{.662}}\relax
\mov(0,#2.45){\hr*{.531}}\mov(0,#2.5){\hr*{.320}}\relax
\special{em:linewidth \the\linwid*}}}}

\def\vdot{\rlap{\kern-1.9pt\lower1.8pt\hbox{$\scriptstyle\bullet$}}}
\def\vtimes{\rlap{\kern-3pt\lower1.8pt\hbox{$\scriptstyle\times$}}}
\def\vDot{\rlap{\kern-2.3pt\lower2.7pt\hbox{$\bullet$}}}
\def\vTimes{\rlap{\kern-3.6pt\lower2.4pt\hbox{$\times$}}}

\def\arc(#1)[#2,#3]{{\k*=#2\l*=#3\m*=\l*
\advance\m*-6\ifnum\k*>\l*\relax\else
{\rotate(#2)\mov(#1,0){\sm*}}\loop
\ifnum\k*<\m*\advance\k*5{\rotate(\k*)\mov(#1,0){\sl*}}\repeat
{\rotate(#3)\mov(#1,0){\sl*}}\fi}}

\def\dasharc(#1)[#2,#3]{{\k**=#2\n*=#3\advance\n*-1\advance\n*-\k**
\L*=1000sp\L*#1\L* \multiply\L*\n* \multiply\L*\Nhalfperiods
\divide\L*57\N*\L* \divide\N*2000\ifnum\N*=0\N*1\fi
\r*\n*  \divide\r*\N* \ifnum\r*<2\r*2\fi
\m**\r* \divide\m**2 \l**\r* \advance\l**-\m** \N*\n* \divide\N*\r*
\k**\r* \multiply\k**\N* \dn*\n* \advance\dn*-\k** \divide\dn*2\advance\dn*\*one
\r*\l** \divide\r*2\advance\dn*\r* \advance\N*-2\k**#2\relax
\ifnum\l**<6{\rotate(#2)\mov(#1,0){\sm*}}\advance\k**\dn*
{\rotate(\k**)\mov(#1,0){\sl*}}\advance\k**\m**
{\rotate(\k**)\mov(#1,0){\sm*}}\loop
\advance\k**\l**{\rotate(\k**)\mov(#1,0){\sl*}}\advance\k**\m**
{\rotate(\k**)\mov(#1,0){\sm*}}\advance\N*-1\ifnum\N*>0\repeat
{\rotate(#3)\mov(#1,0){\sl*}}\else\advance\k**\dn*
\arc(#1)[#2,\k**]\loop\advance\k**\m** \r*\k**
\advance\k**\l** {\arc(#1)[\r*,\k**]}\relax
\advance\N*-1\ifnum\N*>0\repeat
\advance\k**\m**\arc(#1)[\k**,#3]\fi}}

\def\triangarc#1(#2)[#3,#4]{{\k**=#3\n*=#4\advance\n*-\k**
\L*=1000sp\L*#2\L* \multiply\L*\n* \multiply\L*\Nhalfperiods
\divide\L*57\N*\L* \divide\N*1000\ifnum\N*=0\N*1\fi
\d**=#2\Lengthunit \d*\d** \divide\d*57\multiply\d*\n*
\r*\n*  \divide\r*\N* \ifnum\r*<2\r*2\fi
\m**\r* \divide\m**2 \l**\r* \advance\l**-\m** \N*\n* \divide\N*\r*
\dt*\d* \divide\dt*\N* \dt*.5\dt* \dt*#1\dt*
\divide\dt*1000\multiply\dt*\magnitude
\k**\r* \multiply\k**\N* \dn*\n* \advance\dn*-\k** \divide\dn*2\relax
\r*\l** \divide\r*2\advance\dn*\r* \advance\N*-1\k**#3\relax
{\rotate(#3)\mov(#2,0){\sm*}}\advance\k**\dn*
{\rotate(\k**)\mov(#2,0){\sl*}}\advance\k**-\m**\advance\l**\m**\loop\dt*-\dt*
\d*\d** \advance\d*\dt*
\advance\k**\l**{\rotate(\k**)\rmov*(\d*,0pt){\sl*}}%
\advance\N*-1\ifnum\N*>0\repeat\advance\k**\m**
{\rotate(\k**)\mov(#2,0){\sl*}}{\rotate(#4)\mov(#2,0){\sl*}}}}

\def\wavearc#1(#2)[#3,#4]{{\k**=#3\n*=#4\advance\n*-\k**
\L*=4000sp\L*#2\L* \multiply\L*\n* \multiply\L*\Nhalfperiods
\divide\L*57\N*\L* \divide\N*1000\ifnum\N*=0\N*1\fi
\d**=#2\Lengthunit \d*\d** \divide\d*57\multiply\d*\n*
\r*\n*  \divide\r*\N* \ifnum\r*=0\r*1\fi
\m**\r* \divide\m**2 \l**\r* \advance\l**-\m** \N*\n* \divide\N*\r*
\dt*\d* \divide\dt*\N* \dt*.7\dt* \dt*#1\dt*
\divide\dt*1000\multiply\dt*\magnitude
\k**\r* \multiply\k**\N* \dn*\n* \advance\dn*-\k** \divide\dn*2\relax
\divide\N*4\advance\N*-1\k**#3\relax
{\rotate(#3)\mov(#2,0){\sm*}}\advance\k**\dn*
{\rotate(\k**)\mov(#2,0){\sl*}}\advance\k**-\m**\advance\l**\m**\loop\dt*-\dt*
\d*\d** \advance\d*\dt* \dd*\d** \advance\dd*1.3\dt*
\advance\k**\r*{\rotate(\k**)\rmov*(\d*,0pt){\sl*}}\relax
\advance\k**\r*{\rotate(\k**)\rmov*(\dd*,0pt){\sl*}}\relax
\advance\k**\r*{\rotate(\k**)\rmov*(\d*,0pt){\sl*}}\relax
\advance\k**\r*
\advance\N*-1\ifnum\N*>0\repeat\advance\k**\m**
{\rotate(\k**)\mov(#2,0){\sl*}}{\rotate(#4)\mov(#2,0){\sl*}}}}

\def\gmov*#1(#2,#3)#4{\rlap{\L*=#1\Lengthunit
\xL*=#2\L* \yL*=#3\L*
\rx* \gcos*\xL* \tmp* \gsin*\yL* \advance\rx*-\tmp*
\ry* \gcos*\yL* \tmp* \gsin*\xL* \advance\ry*\tmp*
\rx*=\xscale\rx* \ry*=\yscale\ry*
\xL* \the\cos*\rx* \tmp* \the\sin*\ry* \advance\xL*-\tmp*
\yL* \the\cos*\ry* \tmp* \the\sin*\rx* \advance\yL*\tmp*
\kern\xL*\raise\yL*\hbox{#4}}}

\def\rgmov*(#1,#2)#3{\rlap{\xL*#1\yL*#2\relax
\rx* \gcos*\xL* \tmp* \gsin*\yL* \advance\rx*-\tmp*
\ry* \gcos*\yL* \tmp* \gsin*\xL* \advance\ry*\tmp*
\rx*=\xscale\rx* \ry*=\yscale\ry*
\xL* \the\cos*\rx* \tmp* \the\sin*\ry* \advance\xL*-\tmp*
\yL* \the\cos*\ry* \tmp* \the\sin*\rx* \advance\yL*\tmp*
\kern\xL*\raise\yL*\hbox{#3}}}

\def\Earc(#1)[#2,#3][#4,#5]{{\k*=#2\l*=#3\m*=\l*
\advance\m*-6\ifnum\k*>\l*\relax\else\def\xscale{#4}\def\yscale{#5}\relax
{\angle**0\rotate(#2)}\gmov*(#1,0){\sm*}\loop
\ifnum\k*<\m*\advance\k*5\relax
{\angle**0\rotate(\k*)}\gmov*(#1,0){\sl*}\repeat
{\angle**0\rotate(#3)}\gmov*(#1,0){\sl*}\relax
\def\xscale{1}\def\yscale{1}\fi}}

\def\dashEarc(#1)[#2,#3][#4,#5]{{\k**=#2\n*=#3\advance\n*-1\advance\n*-\k**
\L*=1000sp\L*#1\L* \multiply\L*\n* \multiply\L*\Nhalfperiods
\divide\L*57\N*\L* \divide\N*2000\ifnum\N*=0\N*1\fi
\r*\n*  \divide\r*\N* \ifnum\r*<2\r*2\fi
\m**\r* \divide\m**2 \l**\r* \advance\l**-\m** \N*\n* \divide\N*\r*
\k**\r*\multiply\k**\N* \dn*\n* \advance\dn*-\k** \divide\dn*2\advance\dn*\*one
\r*\l** \divide\r*2\advance\dn*\r* \advance\N*-2\k**#2\relax
\ifnum\l**<6\def\xscale{#4}\def\yscale{#5}\relax
{\angle**0\rotate(#2)}\gmov*(#1,0){\sm*}\advance\k**\dn*
{\angle**0\rotate(\k**)}\gmov*(#1,0){\sl*}\advance\k**\m**
{\angle**0\rotate(\k**)}\gmov*(#1,0){\sm*}\loop
\advance\k**\l**{\angle**0\rotate(\k**)}\gmov*(#1,0){\sl*}\advance\k**\m**
{\angle**0\rotate(\k**)}\gmov*(#1,0){\sm*}\advance\N*-1\ifnum\N*>0\repeat
{\angle**0\rotate(#3)}\gmov*(#1,0){\sl*}\def\xscale{1}\def\yscale{1}\else
\advance\k**\dn* \Earc(#1)[#2,\k**][#4,#5]\loop\advance\k**\m** \r*\k**
\advance\k**\l** {\Earc(#1)[\r*,\k**][#4,#5]}\relax
\advance\N*-1\ifnum\N*>0\repeat
\advance\k**\m**\Earc(#1)[\k**,#3][#4,#5]\fi}}

\def\triangEarc#1(#2)[#3,#4][#5,#6]{{\k**=#3\n*=#4\advance\n*-\k**
\L*=1000sp\L*#2\L* \multiply\L*\n* \multiply\L*\Nhalfperiods
\divide\L*57\N*\L* \divide\N*1000\ifnum\N*=0\N*1\fi
\d**=#2\Lengthunit \d*\d** \divide\d*57\multiply\d*\n*
\r*\n*  \divide\r*\N* \ifnum\r*<2\r*2\fi
\m**\r* \divide\m**2 \l**\r* \advance\l**-\m** \N*\n* \divide\N*\r*
\dt*\d* \divide\dt*\N* \dt*.5\dt* \dt*#1\dt*
\divide\dt*1000\multiply\dt*\magnitude
\k**\r* \multiply\k**\N* \dn*\n* \advance\dn*-\k** \divide\dn*2\relax
\r*\l** \divide\r*2\advance\dn*\r* \advance\N*-1\k**#3\relax
\def\xscale{#5}\def\yscale{#6}\relax
{\angle**0\rotate(#3)}\gmov*(#2,0){\sm*}\advance\k**\dn*
{\angle**0\rotate(\k**)}\gmov*(#2,0){\sl*}\advance\k**-\m**
\advance\l**\m**\loop\dt*-\dt* \d*\d** \advance\d*\dt*
\advance\k**\l**{\angle**0\rotate(\k**)}\rgmov*(\d*,0pt){\sl*}\relax
\advance\N*-1\ifnum\N*>0\repeat\advance\k**\m**
{\angle**0\rotate(\k**)}\gmov*(#2,0){\sl*}\relax
{\angle**0\rotate(#4)}\gmov*(#2,0){\sl*}\def\xscale{1}\def\yscale{1}}}

\def\waveEarc#1(#2)[#3,#4][#5,#6]{{\k**=#3\n*=#4\advance\n*-\k**
\L*=4000sp\L*#2\L* \multiply\L*\n* \multiply\L*\Nhalfperiods
\divide\L*57\N*\L* \divide\N*1000\ifnum\N*=0\N*1\fi
\d**=#2\Lengthunit \d*\d** \divide\d*57\multiply\d*\n*
\r*\n*  \divide\r*\N* \ifnum\r*=0\r*1\fi
\m**\r* \divide\m**2 \l**\r* \advance\l**-\m** \N*\n* \divide\N*\r*
\dt*\d* \divide\dt*\N* \dt*.7\dt* \dt*#1\dt*
\divide\dt*1000\multiply\dt*\magnitude
\k**\r* \multiply\k**\N* \dn*\n* \advance\dn*-\k** \divide\dn*2\relax
\divide\N*4\advance\N*-1\k**#3\def\xscale{#5}\def\yscale{#6}\relax
{\angle**0\rotate(#3)}\gmov*(#2,0){\sm*}\advance\k**\dn*
{\angle**0\rotate(\k**)}\gmov*(#2,0){\sl*}\advance\k**-\m**
\advance\l**\m**\loop\dt*-\dt*
\d*\d** \advance\d*\dt* \dd*\d** \advance\dd*1.3\dt*
\advance\k**\r*{\angle**0\rotate(\k**)}\rgmov*(\d*,0pt){\sl*}\relax
\advance\k**\r*{\angle**0\rotate(\k**)}\rgmov*(\dd*,0pt){\sl*}\relax
\advance\k**\r*{\angle**0\rotate(\k**)}\rgmov*(\d*,0pt){\sl*}\relax
\advance\k**\r*
\advance\N*-1\ifnum\N*>0\repeat\advance\k**\m**
{\angle**0\rotate(\k**)}\gmov*(#2,0){\sl*}\relax
{\angle**0\rotate(#4)}\gmov*(#2,0){\sl*}\def\xscale{1}\def\yscale{1}}}

\newcount\CatcodeOfAtSign
\CatcodeOfAtSign=\the\catcode`\@
\catcode`\@=11
\def\@arc#1[#2][#3]{\rlap{\Lengthunit=#1\Lengthunit
\sm*\l*arc(#2.1914,#3.0381)[#2][#3]\relax
\mov(#2.1914,#3.0381){\l*arc(#2.1622,#3.1084)[#2][#3]}\relax
\mov(#2.3536,#3.1465){\l*arc(#2.1084,#3.1622)[#2][#3]}\relax
\mov(#2.4619,#3.3086){\l*arc(#2.0381,#3.1914)[#2][#3]}}}

\def\dash@arc#1[#2][#3]{\rlap{\Lengthunit=#1\Lengthunit
\d*arc(#2.1914,#3.0381)[#2][#3]\relax
\mov(#2.1914,#3.0381){\d*arc(#2.1622,#3.1084)[#2][#3]}\relax
\mov(#2.3536,#3.1465){\d*arc(#2.1084,#3.1622)[#2][#3]}\relax
\mov(#2.4619,#3.3086){\d*arc(#2.0381,#3.1914)[#2][#3]}}}

\def\wave@arc#1[#2][#3]{\rlap{\Lengthunit=#1\Lengthunit
\w*lin(#2.1914,#3.0381)\relax
\mov(#2.1914,#3.0381){\w*lin(#2.1622,#3.1084)}\relax
\mov(#2.3536,#3.1465){\w*lin(#2.1084,#3.1622)}\relax
\mov(#2.4619,#3.3086){\w*lin(#2.0381,#3.1914)}}}

\def\bezier#1(#2,#3)(#4,#5)(#6,#7){\N*#1\l*\N* \advance\l*\*one
\d* #4\Lengthunit \advance\d* -#2\Lengthunit \multiply\d* \*two
\b* #6\Lengthunit \advance\b* -#2\Lengthunit
\advance\b*-\d* \divide\b*\N*
\d** #5\Lengthunit \advance\d** -#3\Lengthunit \multiply\d** \*two
\b** #7\Lengthunit \advance\b** -#3\Lengthunit
\advance\b** -\d** \divide\b**\N*
\mov(#2,#3){\sm*{\loop\ifnum\m*<\l*
\a*\m*\b* \advance\a*\d* \divide\a*\N* \multiply\a*\m*
\a**\m*\b** \advance\a**\d** \divide\a**\N* \multiply\a**\m*
\rmov*(\a*,\a**){\unhcopy\spl*}\advance\m*\*one\repeat}}}

\catcode`\*=12

\newcount\n@ast
\def\n@ast@#1{\n@ast0\relax\get@ast@#1\end}
\def\get@ast@#1{\ifx#1\end\let\next\relax\else
\ifx#1*\advance\n@ast1\fi\let\next\get@ast@\fi\next}

\newif\if@up \newif\if@dwn
\def\up@down@#1{\@upfalse\@dwnfalse
\if#1u\@uptrue\fi\if#1U\@uptrue\fi\if#1+\@uptrue\fi
\if#1d\@dwntrue\fi\if#1D\@dwntrue\fi\if#1-\@dwntrue\fi}

\def\halfcirc#1(#2)[#3]{{\Lengthunit=#2\Lengthunit\up@down@{#3}\relax
\if@up\mov(0,.5){\@arc[-][-]\@arc[+][-]}\fi
\if@dwn\mov(0,-.5){\@arc[-][+]\@arc[+][+]}\fi
\def\lft{\mov(0,.5){\@arc[-][-]}\mov(0,-.5){\@arc[-][+]}}\relax
\def\rght{\mov(0,.5){\@arc[+][-]}\mov(0,-.5){\@arc[+][+]}}\relax
\if#3l\lft\fi\if#3L\lft\fi\if#3r\rght\fi\if#3R\rght\fi
\n@ast@{#1}\relax
\ifnum\n@ast>0\if@up\shade[+]\fi\if@dwn\shade[-]\fi\fi
\ifnum\n@ast>1\if@up\dshade[+]\fi\if@dwn\dshade[-]\fi\fi}}

\def\halfdashcirc(#1)[#2]{{\Lengthunit=#1\Lengthunit\up@down@{#2}\relax
\if@up\mov(0,.5){\dash@arc[-][-]\dash@arc[+][-]}\fi
\if@dwn\mov(0,-.5){\dash@arc[-][+]\dash@arc[+][+]}\fi
\def\lft{\mov(0,.5){\dash@arc[-][-]}\mov(0,-.5){\dash@arc[-][+]}}\relax
\def\rght{\mov(0,.5){\dash@arc[+][-]}\mov(0,-.5){\dash@arc[+][+]}}\relax
\if#2l\lft\fi\if#2L\lft\fi\if#2r\rght\fi\if#2R\rght\fi}}

\def\halfwavecirc(#1)[#2]{{\Lengthunit=#1\Lengthunit\up@down@{#2}\relax
\if@up\mov(0,.5){\wave@arc[-][-]\wave@arc[+][-]}\fi
\if@dwn\mov(0,-.5){\wave@arc[-][+]\wave@arc[+][+]}\fi
\def\lft{\mov(0,.5){\wave@arc[-][-]}\mov(0,-.5){\wave@arc[-][+]}}\relax
\def\rght{\mov(0,.5){\wave@arc[+][-]}\mov(0,-.5){\wave@arc[+][+]}}\relax
\if#2l\lft\fi\if#2L\lft\fi\if#2r\rght\fi\if#2R\rght\fi}}

\catcode`\*=11

\def\Circle#1(#2){\halfcirc#1(#2)[u]\halfcirc#1(#2)[d]\n@ast@{#1}\relax
\ifnum\n@ast>0\L*=\xscale\Lengthunit
\ifnum\angle**=0\clap{\vrule width#2\L* height.1pt}\else
\L*=#2\L*\L*=.5\L*\special{em:linewidth .001pt}\relax
\rmov*(-\L*,0pt){\sm*}\rmov*(\L*,0pt){\sl*}\relax
\special{em:linewidth \the\linwid*}\fi\fi}

\catcode`\*=12

\def\wavecirc(#1){\halfwavecirc(#1)[u]\halfwavecirc(#1)[d]}

\def\dashcirc(#1){\halfdashcirc(#1)[u]\halfdashcirc(#1)[d]}

\def\xscale{1}
\def\yscale{1}

\def\Ellipse#1(#2)[#3,#4]{\def\xscale{#3}\def\yscale{#4}\relax
\Circle#1(#2)\def\xscale{1}\def\yscale{1}}

\def\dashEllipse(#1)[#2,#3]{\def\xscale{#2}\def\yscale{#3}\relax
\dashcirc(#1)\def\xscale{1}\def\yscale{1}}

\def\waveEllipse(#1)[#2,#3]{\def\xscale{#2}\def\yscale{#3}\relax
\wavecirc(#1)\def\xscale{1}\def\yscale{1}}

\def\halfEllipse#1(#2)[#3][#4,#5]{\def\xscale{#4}\def\yscale{#5}\relax
\halfcirc#1(#2)[#3]\def\xscale{1}\def\yscale{1}}

\def\halfdashEllipse(#1)[#2][#3,#4]{\def\xscale{#3}\def\yscale{#4}\relax
\halfdashcirc(#1)[#2]\def\xscale{1}\def\yscale{1}}

\def\halfwaveEllipse(#1)[#2][#3,#4]{\def\xscale{#3}\def\yscale{#4}\relax
\halfwavecirc(#1)[#2]\def\xscale{1}\def\yscale{1}}

\catcode`\@=\the\CatcodeOfAtSign

\newpage
The study of electromagnetic properties of hydrogen-like atoms
and ions in quantum electrodynamics (QED) is one of the basic
tasks in the theory of two-particle bound states. The experimental
verification of the g-factors calculation for the bound particles is
carried out during many years \cite{MT,BT}. The values of the electron
g-factors in hydrogen atom, deuterium, tritium and helium ($^4He^{+}$),
measured in many experiments, are in good agreement with theoretical
results. Recently the scope of the experimental investigation
of hydrogenic ions was essentially inlarged \cite{H,Q}.
These experiments generate a need for the new theoretical study of
different contributions to the gyromagnetic factors of the bound
particles \cite{VS,K,CMY}.
At present, the most accurate value for the electron g-factor
is obtained in the experiment with the hydrogenic carbon ion
${\rm ^{12}C^{5+}}$ (Z=6) \cite{H,Q}:
\begin{equation}
g_e^{exp}(^{12}C^{5+})=2.001~041~596~4(8)(6)(40),
\end{equation}
where the statistical (8), systematical (6) errors and inaccuracy,
connected with the electron mass (40), are shown in brackets.
Theoretical investigations of the electromagnetic properties
of the hydrogen-like atoms, which were done in [8-14], showed that the
gyromagnetic factors of the bound particles can be written in the from:
\begin{equation}
g(H-{\rm atom})=2+\Delta g_{rel}+\Delta g_{rad}+\Delta g_{rec}+...
\end{equation}
Relativistic corrections $\Delta g_{rel}$, radiative corrections $\Delta g_{rad}$,
recoil corrections $\Delta g_{rec}$ in (2) were calculated with the accuracy
up to terms of order $\alpha^3(m/M)$ and $\alpha^2(m/M)^2$ in \cite{F1,F2}
on the basis of the quasipotential method for spin 1/2 particles, composing the
bound system. The dots designate other possible contributions to the g-factor.
In addition, due to the experiments with deuterium, hydrogen-like ions,
which have the nucleus of arbitrary spin, there is need to extend the
calculational methods of the g-factors on this case. In paper
\cite{EG} it was suggested an approach for the calculation of the corrections
to the gyromagnetic factors, based on the Bargmann-Michel-Telegdi (BMT) equation
\cite{t4}. The conclusion about independence of the binding corrections
on the magnitude of the spin of the constituents was also formulated here.
In this work we extended the quasipotential method for the calculation
of the magnetic moment of two-particle bound state to the case of arbitrary
spin particles and calculated main contributions to the corrections in eq.(2).

The interaction of the massive particles of arbitrary spin with the
electromagnetic field is studied on the basis of different methods during
a long time [17-25], but still this problem is far from its final solution.
It was shown in \cite{FPT} that the particle of arbitrary spin must have
at tree level approximation the gyromagnetic factor g=2. In general case,
the matrix element of the electromagnetic current for the particle of arbitrary spin s
is determined by means of (2s+1) form factors (charge, magnetic,
quadruple, et al.). When studied the magnetic moment of the simple atomic
systems, it may be possible to take into account the form factors of the
minimal multipolarity, describing the distributions of the electric charge and
magnetic moment. The one-particle matrix element $J_\mu$ of the electromagnetic current
operator between states with momenta p and q can be written as follows:
\begin{equation}
J_{\mu}=\bar U(p)\left\{\Gamma_\mu F^D_1+\frac{1}{2m}\Sigma_{\mu\nu}
k^\nu F^P_2\right\}U(q)
\end{equation}
The wave function $U(p)$ of particle with arbitrary spin, entering in (3),
can be presented in the form (see, for instance \cite{KMS,KP}):
\begin{equation}
U=\left(\xi\atop\eta\right)=\left(\xi^{\alpha_1\alpha_2...\alpha_p}_{\dot\beta_1
\dot\beta_2...\dot\beta_q}\atop\eta^{\beta_1\beta_2...\beta_q}_{\dot\alpha_1\dot
\alpha_2...\dot\alpha_p}\right),~~~p+q=2S,
\end{equation}
where spin-tensors $\xi$, $\eta$ are symmetrical in upper and lower indexes.
For the particle of half-integer spin p=S+1/2, q=S-1/2. In the case of integer
spin p=q=S. The Lorentz transformation of the spinors $\xi$ and $\eta$
can be written in the form \cite{KP,RF}:
\begin{equation}
\xi=e^{\frac{\vec\Sigma\vec\phi}{2}}\xi_0,~~~\eta=e^{-\frac{\vec\Sigma\vec
\phi}{2}}\xi_0,
\end{equation}
where the direction of the vector $\vec\phi$ coincides with the velocity of
the particle, $th\phi=v$. The generator of the Lorentz transformation
$\vec\Sigma$ is equal:
\begin{equation}
\vec\Sigma=\sum_{i=1}^p\vec\sigma_i-\sum_{i=p+1}^{p+q}\vec\sigma_i,
\end{equation}
and $\vec\sigma_i$ acts on the ith index of the spinor $\xi_0$ as follows:
\begin{equation}
\vec\sigma_i\xi_0=(\vec\sigma_i)_{\alpha_i\beta_i}(\xi_0)_{...\beta_i...}.
\end{equation}
In the standard representation, which is introduced in analogy with the spin
1/2, the free particle wave function (4) may be written with the accuracy
$(v/c)^2$ in the form:
\begin{equation}
U(p)=\left(\left[1+\frac{(\vec\Sigma\vec p)^2}{8m^2}\right]\xi_0  \atop
\frac{\vec\Sigma\vec p}{2m}\xi_0\right)
\end{equation}
The components of the matrix $\Sigma_{\mu\nu}$ in (3) are the generators of
the boosts and rotations \cite{KP,RF}:
\begin{equation}
\Sigma_{n0}=\left(\begin{array}{cc}
\Sigma_n & 0\\
0 & -\Sigma_n\end{array}\right),~~~\Sigma_{mn}=-2i\epsilon_{mnk}
\left(\begin{array}{cc}
s_k  &  0\\
0  &  s_k\end{array}\right),~~\vec s=\frac{1}{2}\sum_{i=1}^{2S}\vec\sigma_i.
\end{equation}
The general expression for the magnetic moment of the two-particle bound
system reads as \cite{F1,F2}:
\begin{equation}
\vec {\cal M}=-\frac{i}{2}\left[\frac{\partial}{\partial\vec\Delta}\times<\vec K_A|\vec J(0)|\vec K_B>\right],
~~\vec\Delta=\vec K_A-\vec K_B,
\end{equation}
where the matrix element of the current operator between the bound states
\begin{equation}
<\vec K_A|J_\mu(0)|\vec K_B>=\int\frac{d\vec p_1 d\vec p_2}{(2\pi)^3}\delta
(\vec p_1+\vec p_2-\vec K_A)\Psi^\ast_{\vec K_A}(\vec p)\Gamma_\mu(\vec p,
\vec q,E_A,E_B)\times
\end{equation}
\begin{displaymath}
\times\Psi_{\vec K_B}(\vec q)\delta(\vec q_1+\vec q_2-\vec K_B)\frac{d\vec q_1 d\vec q_2}{(2\pi)^3},
\end{displaymath}
is expressed by means of the wave function of the bound state $\Psi_{\vec K_B}(\vec p)$
and the generalized vertex function $\Gamma_\mu$ presented in Figure 1.
The vertex function $\Gamma_\mu$ is determined through the five point function:
\begin{equation}
R_\mu=<0|\psi_1(t,\vec x_1)\psi_2(t,\vec x_2)J_\mu(0)\bar\psi_1(\tau,\vec y_1)
\bar\psi_2(\tau,\vec y_2)|0>,
\end{equation}
projected onto the positive-energy states,
\begin{equation}
\Gamma_\mu=G^{-1}R_\mu^{(+)} G^{-1},~~R_\mu^{(+)}=U_1^\ast U_2^\ast R_\mu U_1 U_2,
\end{equation}
where G is the two-particle Green function.
We study the loosely bound composite system, so it is possible to make the perturbation expansion
of all quantities $\Gamma$, R and $G^{-1}$ in particle interaction.
\begin{equation}
\Gamma=\Gamma^{(0)}+\Gamma^{(1)}+...,~~R=R_0+R_1+...,~~G^{-1}=G_0^{-1}-V_1-...~~,
\end{equation}
\begin{eqnarray}
\Gamma^{(0)}&=&G_0^{-1}R_0G_0^{-1},\\
\Gamma^{(1)}&=&G_0^{-1}R_1G_0^{-1}-V_1G_0\Gamma^{(0)}-\Gamma^{(0)}G_0V_1, ...~,
\end{eqnarray}
where $G_0$ is the Green function of two noninteracting particles and $V_1$ is the one-photon
exchange quasipotential (see eq. (19)).
\begin{figure}
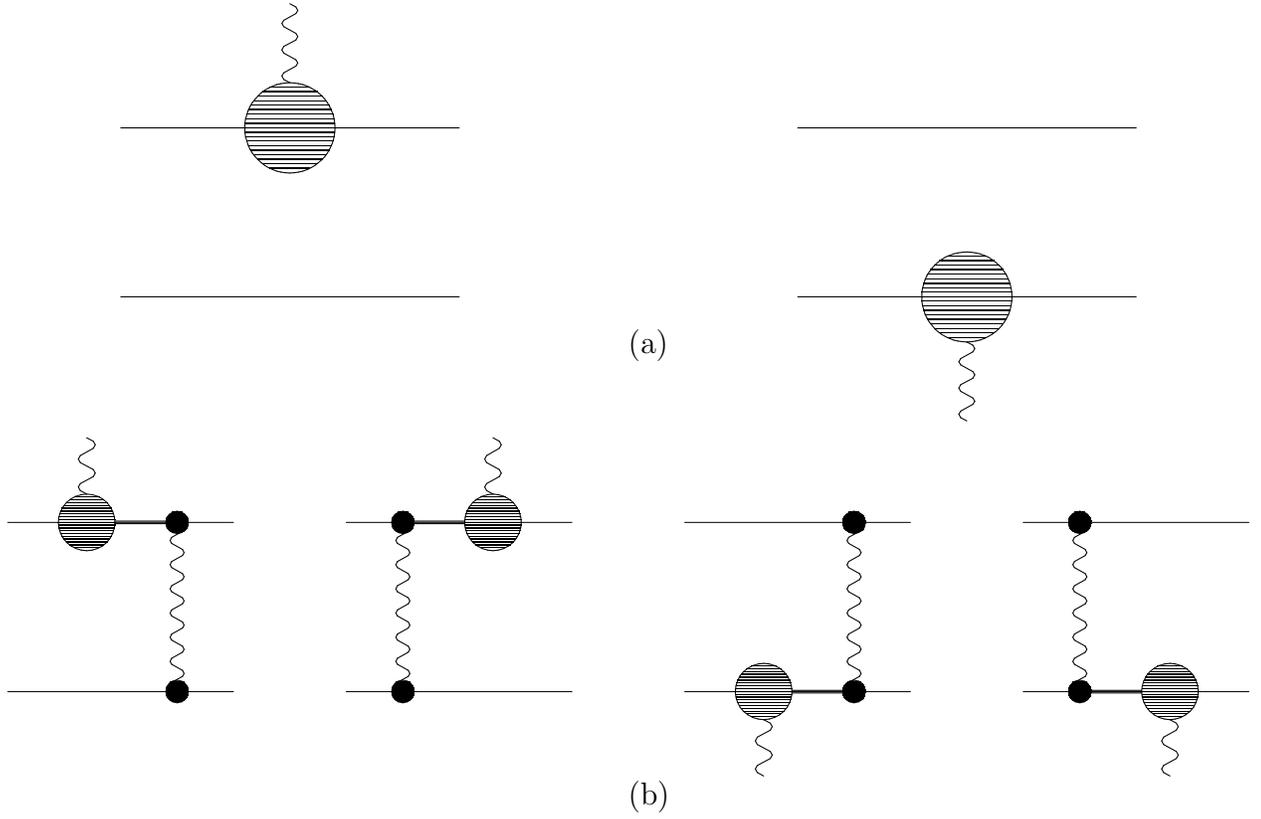

\magnitude=2000
\GRAPH(hsize=15){
\mov(0,0){\lin(2,0)}%
\mov(3,0){\lin(2,0)}%
\mov(6,0){\lin(2,0)}%
\mov(9,0){\lin(2,0)}%
\mov(0,1.5){\lin(2,0)}%
\mov(3,1.5){\lin(2,0)}%
\mov(6,1.5){\lin(2,0)}%
\mov(9,1.5){\lin(2,0)}%
\mov(0.7,1.5){\Circle*(0.5)}%
\mov(1.5,1.5){\wavelin(0,-1.5)}%
\mov(1.5,1.5){\Circle*(0.2)}%
\mov(1.5,1.51){\lin(-0.55,0)}%
\mov(1.5,1.49){\lin(-0.55,0)}%
\mov(3.5,1.5){\Circle*(0.2)}%
\mov(3.5,1.51){\lin(0.55,0)}%
\mov(3.5,1.49){\lin(0.55,0)}%
\mov(7.5,1.5){\Circle*(0.2)}%
\mov(7.5,0.01){\lin(-0.55,0)}%
\mov(7.5,-0.01){\lin(-0.55,0)}%
\mov(9.5,1.5){\Circle*(0.2)}%
\mov(9.5,0.01){\lin(0.55,0)}%
\mov(9.5,-0.01){\lin(0.55,0)}%
\mov(1.5,0.){\Circle*(0.2)}%
\mov(3.5,0.){\Circle*(0.2)}%
\mov(7.5,0.){\Circle*(0.2)}%
\mov(9.5,0.){\Circle*(0.2)}%
\mov(5.5,-1.){(b)}%
\mov(5.5,3.){(a)}%
\mov(0.7,1.75){\wavelin(0,0.5)}%
\mov(3.5,1.5){\wavelin(0,-1.5)}%
\mov(7.5,1.5){\wavelin(0,-1.5)}%
\mov(9.5,1.5){\wavelin(0,-1.5)}%
\mov(4.3,1.5){\Circle*(0.5)}%
\mov(4.3,1.75){\wavelin(0,0.5)}%
\mov(6.7,0.){\Circle*(0.5)}%
\mov(6.7,-0.25){\wavelin(0,-0.5)}%
\mov(10.3,0.){\Circle*(0.5)}%
\mov(10.3,-0.25){\wavelin(0,-0.5)}%
\mov(1.,3.5){\lin(3.,0.)}%
\mov(1.,5.){\lin(3.,0.)}%
\mov(10.,3.5){\lin(-3.,0.)}%
\mov(10.,5.){\lin(-3.,0.)}%
\mov(2.5,5.){\Circle*(0.8)}%
\mov(2.5,5.4){\wavelin(0.,0.7)}%
\mov(8.5,3.5){\Circle*(0.8)}%
\mov(8.5,3.1){\wavelin(0.,-0.7)}%
}
\vspace{3mm}
\caption{Generalized two-particle vertex function $\Gamma_\mu$:
Diagram (a) represents $\Gamma_\mu^{(0)}$, Diagram (b)
describes $\Gamma_\mu^{(1)}$, where the bold line denotes the negative-energy
part of the propagator.}
\end{figure}

The transformation law for the wave function $\Psi_{\vec K_B}(\vec p)$
of the system of bound particles with spins $s_1$, $s_2$ from the rest frame to the reference
frame, moving with the momentum $\vec K_B$, was obtained in \cite{F3}:
\begin{equation}
\delta(\vec p_1+\vec p_2-\vec K_B)\Psi_{\vec K_B}(\vec p)=
D_1^{S_1}(R_W)D_2^{S_2}(R_W)\sqrt{\frac{\epsilon_1^\circ\epsilon_2^\circ M}
{\epsilon_1\epsilon_2 E}}\Psi_0(\vec {p^\circ})\delta(\vec {p_1^\circ}+
\vec {p_2^\circ}),
\end{equation}
where $D^S(R)$ is the well-known rotation matrix and $R_W$ is the Wigner rotation
associated with the Lorentz transformation $\Lambda_{\vec K_B}$: $(E,\vec K_B)=\Lambda_{\vec K_B}
(M,0)$; $(\epsilon,\vec p)=\Lambda(\epsilon^\circ,\vec {p^\circ})$, $E=\sqrt{M^2+\vec K_B^2}$,
$\epsilon(\vec p)=\sqrt{\vec p^2+m^2}$. The exact expression for this
rotation matrix is the following \cite{F2}:
\begin{equation}
D^S(R^W)=S^{-1}(\vec p)S(\vec K_B)S(\vec {p^\circ}),
\end{equation}
where $S(\vec p)$ is the Lorentz transformation matrix of the spinor
wave function (4).
The quasipotential bound state wave function $\Psi_0(\vec {p^\circ})$ in the rest
frame of the composite system satisfies to the quasipotential equation
\cite{MF}:
\begin{equation}
G^{-1}_0\Psi\equiv\left(\frac{b^2}{2\mu_R}-\frac{\vec {p^\circ}^2}
{2\mu_R}\right)\Psi_0(\vec {p^\circ})=
\int V(\vec {p^\circ},\vec {q^\circ},M)\Psi_0(\vec {q^\circ})\frac{d\vec {q^\circ}}
{(2\pi)^3},
\end{equation}
where the relativistic reduced mass
\begin{displaymath}
\mu_R=\frac{E_1E_2}{M}=\frac{M^4-(m_1^2-m_2^2)^2}{4M^3}, E_{1,2}=\frac{M^2-m_{2,1}^2+m_{1,2}^2}
{2M},
\end{displaymath}
$M=E_1+E_2$ is the bound state mass,
\begin{displaymath}
b^2(M)=\frac{[M^2-(m_1+m_2)^2][M^2-(m_1-m_2)^2]}{4M^2}.
\end{displaymath}
In the nonrelativistic limit this equation reduces to the Shroedinger
equation with the Coulomb potential. The $D^S$ functions in (18)
can be obtained in the approximate form, using (5):
\begin{equation}
D^S(R_W)\approx 1+\frac{{\vec {p^\circ}^2}-(\vec\Sigma\vec {p^\circ})
(\vec\Sigma\vec {p^\circ})}
{4m^2}+\frac{\vec K_B^2-(\vec\Sigma\vec K_B)(\vec\Sigma\vec K_B)}{4M^2}+
\frac{\vec {p^\circ}\vec K_B-(\vec\Sigma\vec {p^\circ})(\vec\Sigma\vec K_B)}{4mM}.
\end{equation}
The main contribution to the vertex function $\Gamma_\mu$, which is determined
by Feynman diagram (a), shown in Figure 1, can be expressed as follows
(in the Breit frame):
\begin{equation}
\vec\Gamma^{(0)}(\vec p,\vec q)=\bar U_1(p_1)e_1\Biggl\{\vec\Gamma_1
+\frac{i\kappa_1}{m_1}\left[\vec S_1\times\vec\Delta\right]\Biggr\}U_1(q_1)
\delta(\vec p_2-\vec q_2)+(1\leftrightarrow 2),
\end{equation}
\begin{displaymath}
\vec S_1=\left(\begin{array}{cc}
\vec s_1& 0\\
0& \vec s_1\end{array}\right),~\vec\Delta=\vec p_1-\vec q_1,
\end{displaymath}
where $F^D_{1,2}(0)=e_{1,2}$, $F^P_{1,2}(0)=e_{1,2}\kappa_{1,2}$, and the matrix
\begin{equation}
\vec\Gamma=\left(\begin{array}{cc}
0 & \vec\Sigma\\
-\vec\Sigma & 0\end{array}\right)
\end{equation}
is the natural generalization of the corresponding expression of the Dirac
matrices $\vec\gamma$ for the spin 1/2 in the standard representation.
To simplify different terms of eq. (21) it is helpful to use the following
relationships \cite{KP}:
\begin{equation}
[\Sigma_i,\Sigma_j]=4i\epsilon_{ijk}s_k,~~~[\Sigma_i,s_j]=i\epsilon_{ijk}\Sigma_k.
\end{equation}
To construct vertex function $\vec\Gamma^{(0)}(\vec p,\vec q)$, accounting for the terms $(v/c)^2$, we use explicit expression of the wave function (8),
transforming the different parts of the matrix element (21) by means of the
equation of motion for the spinor U(p). The following relations are valid
(taking into account $\delta(\vec p_2-\vec q_2)$):
\begin{equation}
\bar U_1(\vec p_1)\frac{\vec p_1+\vec q_1}{2m_1}U(\vec q_1)=2\vec {p^\circ}-
\frac{\epsilon_2}{M}\vec\Delta+\frac{i\vec {p^\circ}\left(\vec S_1[\vec {p^\circ}\times\vec
\Delta]\right)}{m_1^2},
\end{equation}
\begin{equation}
\bar U_1(\vec p_1)\frac{\epsilon_1(\vec p_1)-\epsilon_1(\vec q_1)}{2m_1}
\vec{\cal A}_1U_1(\vec q_1)=
-\frac{2\vec {p^\circ}\vec \Delta}{m_1^2}i[\vec S_1\times\vec {p^\circ}],
\end{equation}
\begin{equation}
\bar U_1(\vec p_1)[\vec S_1\times\vec\Delta]U_1(\vec q_1)=[\vec S_1\times\vec\Delta]-
\frac{1}{2m_1^2}\left\{\vec {p^\circ}\left(\vec S_1[\vec {p^\circ}\times\vec\Delta]\right)+
[\vec {p^\circ}\times\vec S_1](\vec {p^\circ}\vec\Delta)\right\}.
\end{equation}
Bound state effects in the vertex function $\Gamma_\mu$ are determined
by the diagram (b) of Figure 1. Accounting for iteration terms with the quasipotential,
we can represent the corresponding expression in the form \cite{F1,F2}:
\begin{equation}
\vec\Gamma^{(1)}(\vec p,\vec q)=U_1^\ast(\vec p_1)U_2^\ast(\vec p_2)\frac{e_1}
{2m_1}\Biggl\{\vec{\cal A}_1\Lambda_1^{(-)}(\vec p_1'){\cal B}_1{\cal B}_2
\hat V(\vec q_2-\vec p_2)+
\end{equation}
\begin{displaymath}
+{\cal B}_1{\cal B}_2\hat V(\vec p_2-\vec q_2)\Lambda_1^{(-)}(\vec q_1')
\vec{\cal A}_1\Biggr\}U_1(\vec q_1)U_2(\vec q_2)+(1\leftrightarrow 2),
\end{displaymath}
\begin{equation}
\hat V(\vec k)={\cal B}_1{\cal B}_2\Biggl\{\left(1+\frac{\kappa_1}{2m_1}\vec\Gamma_1
\vec k\right)\left(1-\frac{\kappa_2}{2m_2}\vec\Gamma_2\vec k\right)-
\end{equation}
\begin{displaymath}
-\left(\vec{\cal A}_1+\frac{\kappa_1}{m_1}{\cal B}_1i[\vec S_1\times\vec k]\right)
\left(\vec{\cal A}_2-\frac{\kappa_2}{m_2}{\cal B}_2i[\vec S_2\times\vec k]\right)
\Biggr\}\frac{e_1e_2}{k^2},
\end{displaymath}
where the negative-energy projection operator
$\Lambda^-(\vec p)\approx (1-{\cal B})/2-\vec{\cal A}\vec p/2m$, $\kappa_{1,2}$ are
the anomalous magnetic moments of the particles. The matrices $\vec{\cal A}_{1,2}$,
${\cal B}_{1,2}$ are also the generalizations for $\vec\alpha_{1,2}$, $\beta_{1,2}$
used in the case of spin 1/2 particles as in (29):
\begin{equation}
\vec{\cal A}=\left(\begin{array}{cc}
0 & \vec\Sigma\\
\vec\Sigma & 0\end{array}\right),~~
{\cal B}=\left(\begin{array}{cc}
I& 0\\
0& -I\end{array}\right).
\end{equation}
Both parts of the quasipotential (28) give contribution to the magnetic
moment of the system with the accuracy $(v/c)^2$. Substituting
(27), (28) and (8) in (11) and calculating derivative in $\vec\Delta$ to eq. (10),
we obtain:
\begin{equation}
\vec {\cal M}=\frac{1}{(2\pi)^3}\int d\vec p\Psi_0^\ast(\vec p)\frac{e_1}{2\epsilon_1(\vec p)}
\Biggl\{2(1+\kappa_1)\vec s_1[1+N_1+N_2]+(1+4\kappa_1)\frac{\left[\vec
p\times[\vec s_1\times\vec p]\right]}{2m_1^2}+
\end{equation}
\begin{displaymath}
+(1+\kappa_2)\frac{\left[\vec p\times[\vec s_2\times\vec p]\right]}{m_1m_2}
\frac{\vec\Sigma_1^2}{3}-\frac{\epsilon_2(\vec p)}{M}\left[1+N_1+N_2+
\frac{(M-\epsilon_1-\epsilon_2)}{m_2}\frac{\vec\Sigma_1^2}{3}\right]i
\left[\vec p\times\frac{\partial}{\partial\vec p}\right]+
\end{displaymath}
\begin{displaymath}
+\frac{1}{2M}\left[\vec p\times\left[\vec p\times\left(\frac{\vec s_1}{m_1}-
\frac{\vec s_2}{m_2}\right)\right]\right]\Biggr\}\Psi_0(\vec p)+(1\leftrightarrow 2),
\end{displaymath}
where
\begin{equation}
N_i=\frac{\vec p^2-(\vec\Sigma_i\vec p)(\vec\Sigma_i\vec p)}{2m_i^2}.
\end{equation}
In the case of S-states the expression (30) may be essentially simplified:
\begin{equation}
\vec {\cal M}=\frac{1}{2}g_{1~bound}\frac{e_1}{m_1}<\vec s_1>+\frac{1}{2}g_{2~bound}
\frac{e_2}{m_2}<\vec s_2>,
\end{equation}
where the g-factors of the bound particles are equal:
\begin{equation}
g_{1~bound}=g_1\Biggl\{1-\frac{<\vec p^2>}{3m_1^2}\left[1-\frac{3\kappa_1}{2
(1+\kappa_1)}\right]+\frac{<\vec p^2>}{2m_1^2}\Biggl[1-\frac{<\vec
\Sigma_1^2>}{3}+
\end{equation}
\begin{displaymath}
+\frac{m_1^2}{m_2^2}\left(1-\frac{<\vec\Sigma_2^2>}{3}\right)\Biggr]
+\frac{e_2}{e_1}\frac{<\vec p^2>}{3m_2^2}\frac{<\vec\Sigma_2^2>}{3}-
\frac{<\vec p^2>}{(1+\kappa_1)6m_1(m_1+m_2)}\left(1-\frac{e_2}{e_1}
\frac{m_1}{m_2}\right)\Biggr\},
\end{displaymath}
\begin{displaymath}
g_{2~bound}=g_{1~bound}(1\leftrightarrow 2),~~~~~~~~~~~~~~~~~~~~ \frac{1}{2}g_{1,2}=1+\kappa_{1,2}.
\end{displaymath}
For the hydrogen-like ion (1 is the electron, 2 is the nucleus)
we have: $e_1=-e$, $e_2=Ze$, $<\vec p^2>=m_1^2m_2^2
(Z\alpha)^2/(m_1+m_2)^2$,
\begin{equation}
K_{s_1}=\frac{<\vec\Sigma_1^2>}{3}=1,~K_{s_2}=\frac{<\vec\Sigma_2^2>}{3}=
\Biggl\{{\frac{4s_2}{3},~s_2~{\rm is~the~integer~nucleus~spin}\atop
\frac{4s_2+1}{3},~s_2~{\rm is~the~half-integer~nucleus~spin}},
\end{equation}
so the g-factors of the bound electron and nucleus take the form:
\begin{equation}
g_{e~bound}=g_e\Biggl\{1-\frac{m_2^2(Z\alpha)^2}{3(m_1+m_2)^2}\Biggl[
1-\frac{3\kappa_1}{2(1+\kappa_1)}-\frac{3}{2}\left(1-K_{s_1}\right)-
\end{equation}
\begin{displaymath}
-\frac{3}{2}\frac{m_1^2}{m_2^2}\left(1-K_{s_2}-\frac{2}{3}ZK_{s_2}\right)
+\frac{m_1}{2(m_1+m_2)(1+\kappa_1)}\left(1+Z\frac{m_1}{m_2}\right)\Biggr]\Biggr\},
\end{displaymath}
\begin{equation}
g_{N~bound}=g_N\Biggl\{1-\frac{m_1^2(Z\alpha)^2}{3(m_1+m_2)^2}\Biggl[
1-\frac{3\kappa_2}{2(1+\kappa_2)}-\frac{3}{2}\left(1-K_{s_2}\right)-
\end{equation}
\begin{displaymath}
-\frac{3}{2}\frac{m_2^2}{m_1^2}(1-K_{s_1}-\frac{2}{3Z}K_{s_1})
+\frac{m_2}{2(m_1+m_2)(1+\kappa_2)}\left(1+\frac{m_2}{Zm_1}\right)\Biggr]\Biggr\},
\end{displaymath}
The expressions (35)-(36) were obtained from interaction amplitudes, shown
in Figure 1. They contain corrections of order $O(\alpha^2)$, $O(\alpha^3)$,
connected with the bound state effects. As it follows from (30),
some of these contributions can be considered as relativistic corrections
for spin s particles (terms proportional to $N_i$). Other corrections
refer to the spin-orbit interaction. Our calculation of relations (35)-(36)
shows, that the terms of order $O(\alpha^2)$ in $g_{e~bound}$, $g_{N~bound}$
depend on the spin of second particle - nucleus. This conclusion
differs from the results of the paper \cite{EG}, where such dependence is
absent. In the case of spin 1/2 particles the obtained expressions (35) and (36)
for the bound state g-factors coincide with the results of \cite{F1,G,GH,CO}. The electron
g-factors in hydrogen atom, deuterium, tritium as well as their ratios are very
important from the experimental point of view \cite{MT}. Experimental value of the ratio
$g_{e~H}/g_{e~D}$, obtained in
\cite{WPK} with high accuracy
\begin{equation}
r^{exp}=\left[\frac{g_{e~H}}{g_{e~D}}\right]^{exp}=1+7.22(3)\cdot 10^{-9}.
\end{equation}
The theoretical expression for this ratio can be written from (35) as follows:
\begin{equation}
r^{th}=\left[\frac{g_{e~H}}{g_{e~D}}\right]^{th}=1+\alpha^2\left[\frac{1}{4}
\frac{m_1}{m_2}-\frac{25}{72}\frac{m_1^2}{m_2^2}-\frac{\alpha}{\pi}\left(\frac{m_1}
{24m_2}-\frac{1}{16}\frac{m_1^2}{m_2^2}\right)\right].
\end{equation}

As we pointed out above, the new approach to the calculation of the various order
contributions to the magnetic moment of the loosely bound system was
suggested in the work \cite{EG}. This method is based on the relativistic
semiclassical equation of motion for spin. Constructed in \cite{EG}
on the basis of this equation the interaction Hamiltonian of particles of arbitrary spin
with the external electromagnetic field leads to the g-factors
of the bound particles, which are independent of its spin. It is
known that the BMT equation is approximate one: it is linear on the particle
spin, field strength $F_{\mu\nu}$, which doesn't contain coordinate dependence.
When the spin s particle is in the bound state in the external homogeneous
magnetic field, some terms, omitted in the approximation of the BMT equation,
can give definite contribution to the g-factors of the bound particles.
In this work during the calculation of nucleus spin-dependent terms in the
gyromagnetic factors of the bound particles, composing hydrogen-like ion,
we used the approach, proposed in the papers \cite{KMS,KP}, for the description
of the electromagnetic interactions of particles with arbitrary spin. New
contributions to (35), (36), (38), as compared with \cite{F1}, were obtained
after the replacement of the ordinary boost generators $\vec\alpha$ in the case
of spin 1/2 particles by operators (29). The value of the correction in $r^{th}$,
connected with the deuteron spin is equal to
$\Delta r^{th}=5\alpha^2m_1^2/72m_2^2=0.001\cdot 10^{-9}$
(I=1, Z=1, $m_{nucl}=2m_2$ ($m_2$ is the proton mass). It lies within the limits
of the experimental error, as it follows from eq. (37). Numerically, the ratio (38)
$r^{th}=1+7.237\cdot 10^{-9}$, which is in good agreement with (30). Nucleus
spin dependent corrections in (35), (36) are the functions of Z and the number
of nucleons in the nucleus N. Despite the increasing of these corrections with
Z as ${\rm Z^3}$ the growth of the nucleon number N in nucleus leads to opposite
effect. So, in the case of ions with nucleous spin ${\rm I\neq 0}$ the numerical value
of these corrections is out of experimental accuracy reached for $^{12}C^{5+}$.
For the ion ${\rm ^{12}C^{5+}}$ I=0, so ${\rm K_I=0}$ and the corresponding
spin-dependent correction is equal to zero. At present time the
measurements of the electron g-factors in the ions ${\rm ^{16}O^{7+}}$ and
${\rm^{32}S^{14+}}$ were carried out \cite{Q}, but their nucleus have also spin I=0.
From our viewpoint, it will be interesting to measure the values of electron
g-factors, using the Penning traps \cite{H,Q} for such ions, which have nucleous spin
${\rm I\neq 0}$ and the ratio ${\rm Z^3/N^2}$ would be reached large values. One of the
such ions is the ion $^{59}{\rm Co^{26+}}$, which has ${\rm I=7/2}$, ${\rm Z^3/N^2\approx
5.65}$, and the value of spin-dependent correction in (35) is equal to
$0.1\cdot 10^{-9}$. As was pointed out by W.Quint \cite{Q},
the measurement of the bound electron g-factors with the accuracy
exceeding 1 ppb may be realized in the near future.

We are grateful to W.Quint and V.M. Shabaev for the information about
new experimental data on the g-factors of bound particles, and to S.G. Karshenboim,
I.B. Khriplovich
for useful discussion of the questions, connected with the magnetic moment
of the two-particle bound state. The work was performed under the financial
support of the Russian Foundation for Fundamental Research
(grant 00-02-17771) and the Program "Universities of Russia - Fundamental
Researches" (grant 990192).


\begin{thebibliography}{99}
\bibitem{MT}Mohr P.J., Taylor B.N. Reviews of Mod. Phys. 2000. V.72. P.351.
\bibitem{BT}Beier T. Phys. Rep. 2000. V.339. P.79.
\bibitem{H}Hermanspahn N., et al. Phys. Rev. Lett. 2000. V.84. P.427
\bibitem{Q}Quint W. The g-Factor of the Bound Electron in Hydrogen-like Ions:
a High-Accuracy Test of Bound-State QED, invited talk at $2^{\rm nd}$ Workshop
on Frontier Tests of Quantum Electrodynamics and Physics of the Vacuum (QED 2000),
5-11 October 2000, the Abdus Salam ICTP, to be published in Proceedings QED 2000.
\bibitem{VS}Shabaev V.M. Can. J. Phys. 1998. V.76. P.907.
\bibitem{K}Karshenboim S.G. Phys. Lett. 2000. V.A266. P.380.
\bibitem{CMY}Czarnecki A., Melnikov K., Yelkhovsky A. Anomalous magnetic
moment of a bound electron, Preprint SLAC-PUB-8522, hep-ph/0007217.
\bibitem{F1}Faustov R.N. Phys. Lett. 1970. V.B33. P.422.
\bibitem{F2}Faustov R.N. Nuovo Cimento. 1970. V.69. P.37.
\bibitem{G}Grotch H. Phys. Rev. Lett. 1970. V.24. P.39.
\bibitem{GH}Grotch H., Hegstrom R.A. Phys. Rev. 1971. V.A4. P.59.
\bibitem{BCS}Blundell S.A., Cheng K.T., Sapirstein J. Phys. Rev. 1997. V.A55.
P.1857.
\bibitem{WPK}Walther F.G., Phillips W.D., Kleppner D. Phys. Rev. Lett. 1972.
V.28. P.1159.
\bibitem{PSSL}Persson H., Salomonson S., Sunnergren P., Lindgren I.
Phys. Rev. 1997. V.A56. P.R2498.
\bibitem{EG}Eides M.I., Grotch H. Annals of Phys. 1997. V.260. P.191.
\bibitem{t4}Berestetskii V.B., Lifshitz E.M., Pitaevskii L.P. "Quantum
Electrodynamics", M., Nauka, 1989.
\bibitem{Singh}Singh L.P.S., Hagen C.R. Phys. Rev. 1974. V.D9. P.898;
P.910.
\bibitem{FPT}Ferrara S., Poratti M., Telegdi V.L. Phys. Rev. 1992. V.D46. P.3529.
\bibitem{BP}Brodsky S.D., Primack J.R. Ann. Phys. 1970. V.52. P.315.
\bibitem{CO}Close F.E., Osborn H. Phys. Lett. 1971. V.B34. P.400.
\bibitem{FN}Fushich V.I., Nikitin A.G. Symmetry of Quantum Mechanics Equations,
M., Nauka, 1990.
\bibitem{DPW}Deser S., Pascalutsa V., Waldron A. Massive Spin 3/2 Electrodynamics,
Preprint BRX-TH 469; hep-ph/0003011.
\bibitem{Z}Zinoviev Yu.M. Proc. XVII Seminar on High Energy Physics and
Field Theory, Protvino, 1994, P.189.
\bibitem{KMS}Khriplovich I.B., Milstein A.I., Sen'kov R.A. JETF. 1997. V.111.
P.1935.
\bibitem{KP}Khriplovich I.B., Pomeransky A.A. JETF. 1998. V.113. P.1537.
\bibitem{RF}Rumer U.B., Fet A.I. Group theory and Quantum Fields, M., Nauka,
1977.
\bibitem{F3}Faustov R.N. Ann. Phys. 1973. V.78. P.176.
\bibitem{MF}Martynenko A.P., Faustov R.N. Theor. Math. Phys. 1985. V.64. P.765.
\end{thebibliography}
\end{document}